\documentclass[conf]{new-aiaa} 
\usepackage[utf8]{inputenc}

\usepackage{graphicx}
\usepackage{amsmath}
\usepackage[version=4]{mhchem}
\usepackage{siunitx}
\usepackage{longtable,tabularx}
\usepackage{amssymb}
\usepackage{esvect}
\usepackage{subfigure}
\usepackage{psfrag}
\usepackage[percent]{overpic}
\usepackage{color}
\usepackage{placeins}
\usepackage{lineno}
\usepackage{rotating}
\usepackage{multirow} 
\usepackage{booktabs}     
\usepackage{mathtools}
\usepackage{ifthen}
\usepackage{tikz}
\newcommand*\circled[1]{\tikz[baseline=(char.base)]{
            \node[shape=circle,draw,inner sep=1.1pt] (char) {#1};}}
\newcommand*\dft{\mathop{}\!\mathrm{d}}
\usepackage{nomencl}
\usepackage{etoolbox}
\usepackage{xstring}
\newcommand{\nomenclheader}[1]{
\item[\hspace*{-\itemindent}\normalfont\bfseries #1]}
\renewcommand\nomgroup[1]{
        \IfStrEqCase{#1}{
	{A}{\nomenclheader{Acronyms}}
	{D}{\nomenclheader{Dimensional Properties}}
	{N}{\nomenclheader{Non-dimensional Properties}}
	{O}{\nomenclheader{Operators}}
	{S}{\nomenclheader{Subscripts and Superscripts}}
}
}

\makenomenclature

\graphicspath{{figures}}

\title{Revisiting Wells Turbine Hysteresis in Light of Existing Literature on Moving Airfoils}

\author{Tiziano Ghisu\footnote{Assistant Professor}}
\author{Francesco Cambuli\footnote{Assistant Professor}} 
\author{Pierpaolo Puddu\footnote{Full Professor}} 
\author{Irene Virdis\footnote{Research Assistant}}
\author{Mario Carta\footnote{Research Student}}
\author{Fabio Licheri\footnote{Research Assistant}}
\affil{Department of Mechanical, Chemical and Materials Engineering, University of Cagliari, via Marengo 2, 09123 Cagliari, Italy}

\begin{document}

\maketitle

\nomenclature[A]{CFD}{computational fluid dynamics}
\nomenclature[A]{LPM}{lumped parameter model}
\nomenclature[A]{OWC}{oscillating water column}
\nomenclature[Na]{$A,B,C,D$}{coefficients of second order equation}
\nomenclature[Nb]{$c_d$}{drag coefficient}
\nomenclature[Nc]{$c_l$}{lift coefficient}
\nomenclature[Nd]{$c_m$}{pitching moment coefficient}
\nomenclature[Ne]{$c_x$}{turbine axial force coefficient}
\nomenclature[Ne]{$c_{x,\phi}$}{slope of $c_x$ vs. $\phi$ curve}
\nomenclature[Nf]{$G$}{transfer function}
\nomenclature[Ng]{$j$}{imaginary unit}
\nomenclature[Nh]{$k$}{non-dimensional (or reduced) frequency}
\nomenclature[Ni]{$M$}{Mach number}
\nomenclature[Nj]{$P^*$}{pressure drop coefficient}
\nomenclature[Nk]{$Re$}{Reynolds number}
\nomenclature[Nl]{$t^*$}{non-dimensional time}
\nomenclature[Nm]{$T^*$}{torque coefficient}
\nomenclature[Nn]{$\gamma$}{ratio of specific heats}
\nomenclature[No]{$\phi_p$}{piston-based flow coefficient}
\nomenclature[Np]{$\phi_l$}{local flow coefficient}
\nomenclature[Nq]{$\rho^*$}{non-dimensional density}
\nomenclature[Nr]{$\sigma$}{turbine solidity}
\nomenclature[Ns]{$\xi$}{phase shift}
\nomenclature[Nt]{$\zeta$}{damping ratio}
\nomenclature[Da]{$a$}{speed of sound \si{[m\; s^{-1}]}}
\nomenclature[Db]{$A$}{cross area \si{[m^2]}}
\nomenclature[Dc]{$c$}{blade chord \si{[m]}}
\nomenclature[Dd]{$f$}{frequency \si{[s^{-1}]}}
\nomenclature[De]{$F_x$}{turbine axial force \si{[kg\; m\; s^{-2}]}}
\nomenclature[Df]{$L$}{turbine duct length \si{[m]}}
\nomenclature[Dg]{$M_1$}{mass of air in the chamber \si{[kg]}}
\nomenclature[Dh]{$h_1$}{air chamber height \si{[m]}}
\nomenclature[Di]{$p$}{pressure \si{[kg\; m^{-1} s^{-2}]}}
\nomenclature[Dj]{$r_m$}{blade midspan radius \si{[m]}}
\nomenclature[Dk]{$r_t$}{blade tip radius \si{[m]}}
\nomenclature[Dl]{$t$}{time \si{[s]}}
\nomenclature[Dm]{$T$}{turbine torque \si{[kg\; m^2\; s^{-2}]}}
\nomenclature[Dn]{$U$}{blade speed \si{[m\; s^{-1}]}}
\nomenclature[Do]{$U_{\infty}$}{free-stream velocity \si{[m\; s^{-1}]}}
\nomenclature[Dp]{$V$}{axial velocity \si{[m\; s^{-1}]}}
\nomenclature[Dq]{$\Delta p$}{turbine pressure drop \si{[kg\; m^{-1}\; s^{-2}]}}
\nomenclature[Dr]{$\rho$}{air density \si{[kg\; m^{-3}]}}
\nomenclature[Ds]{$\omega$}{turbine rotational speed \si{[s^{-1}]}}
\nomenclature[Dt]{$\Omega$}{piston angular frequency \si{[s^{-1}]}}
\nomenclature[Du]{$\Omega_n$}{angular natural frequency \si{[s^{-1}]}}
\nomenclature[S]{$0$}{amplitude}
\nomenclature[S]{$1$}{air chamber}
\nomenclature[S]{$2$}{turbine duct}
\nomenclature[S]{$a$}{turbine duct outlet section}
\nomenclature[S]{$f$}{turbine duct inlet section}
\nomenclature[S]{$l$}{local}
\nomenclature[S]{$p$}{piston}
\nomenclature[S]{$t$}{tangential direction}
\nomenclature[S]{$x$}{axial direction}

\printnomenclature 

\section{Introduction. Wells turbines and OWC systems}
\label{sec:wells}

\begin{figure}[!ht]
 \centering
 \subfigure[]{\includegraphics[width=.45\columnwidth]{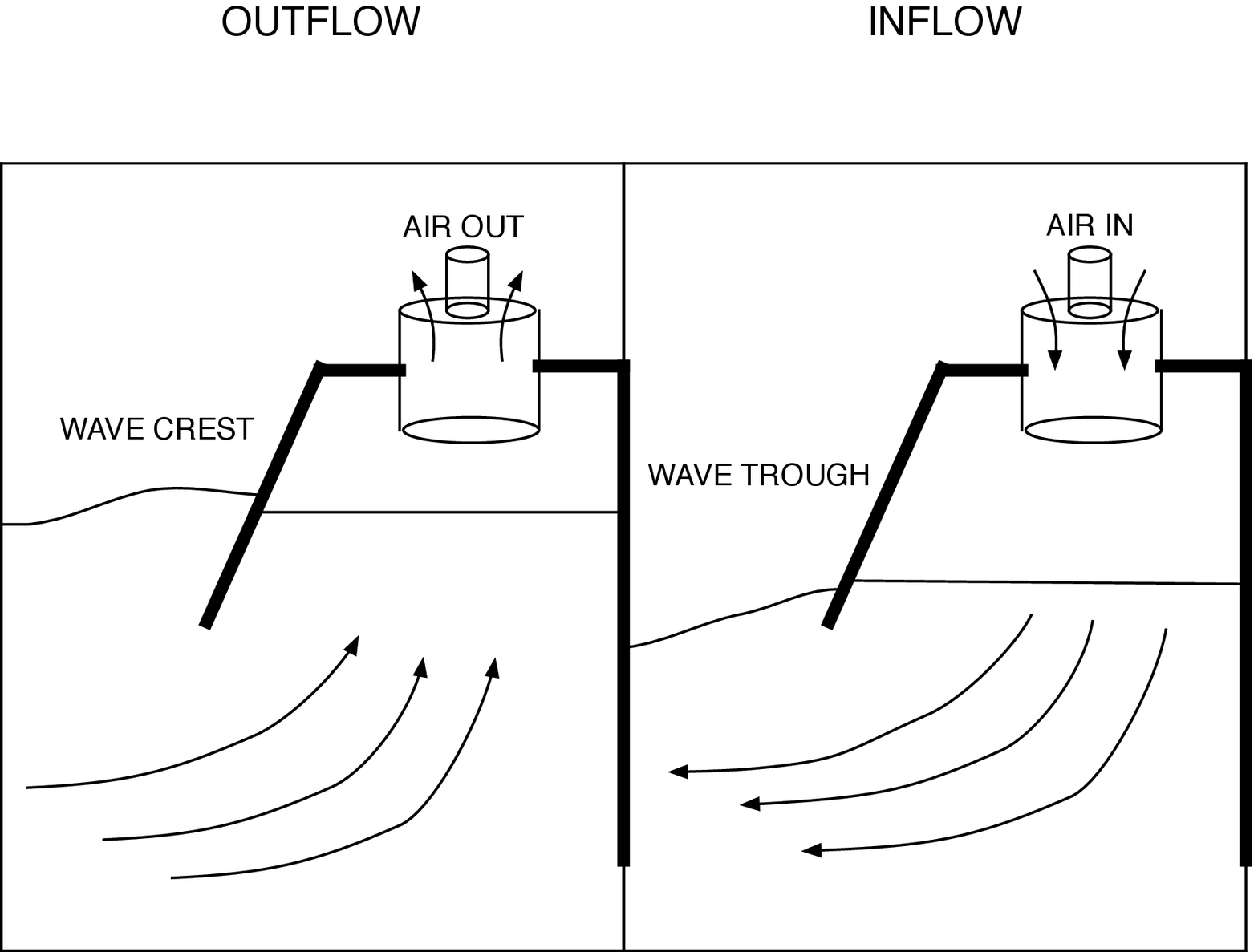}\label{fig:OWC}}
 \hspace{1cm}
 \subfigure[]{\includegraphics[width=.45\columnwidth]{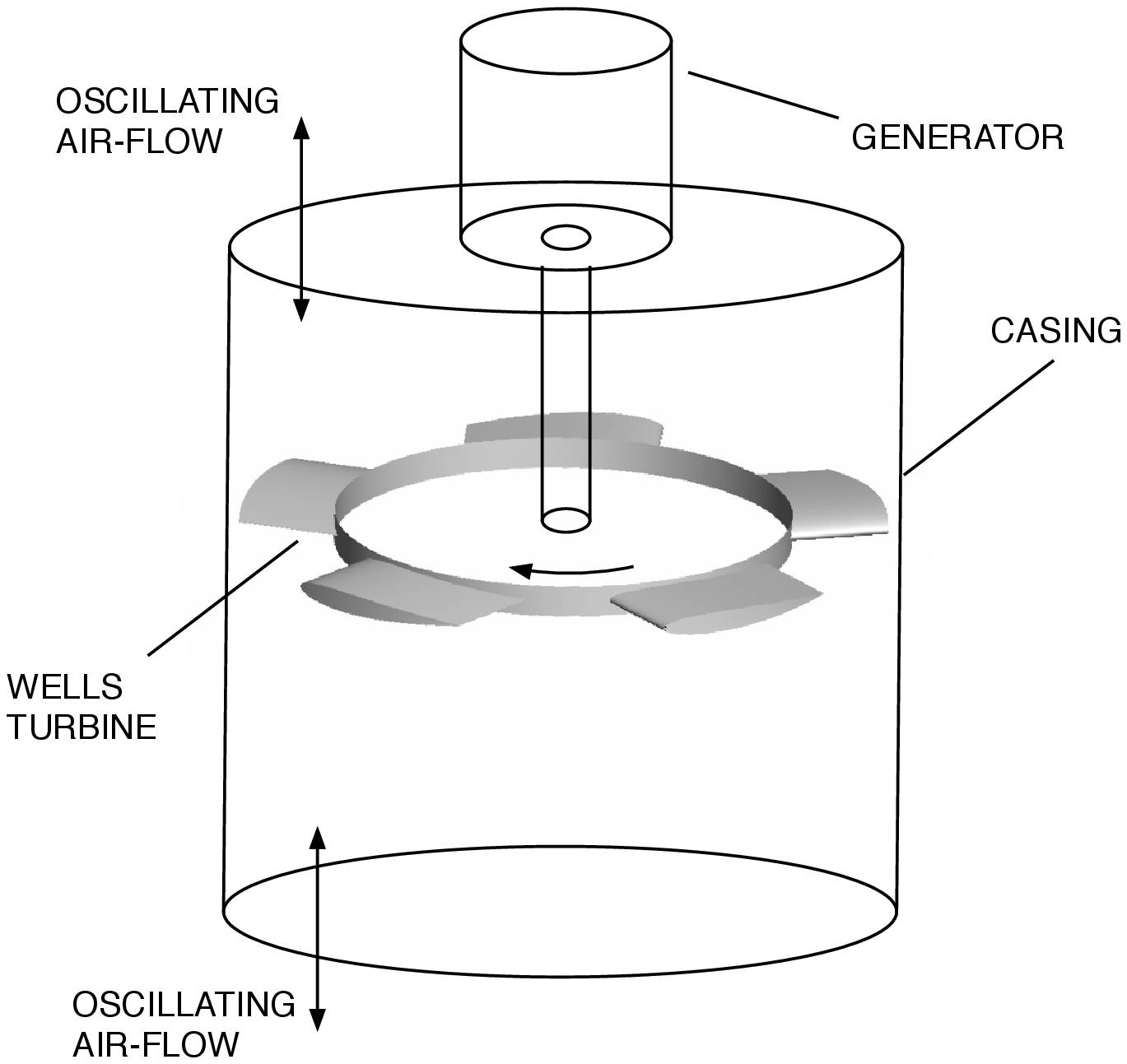}\label{fig:wells}}
\caption{OWC system (a) and Wells turbine (b)}
\label{fig:owc-wells}
\end{figure}

A Wells turbine is an axial-flow turbine consisting of a rotor usually with symmetric (uncambered) blades staggered at a 90 degree angle relative to the incoming flow. This turbine is used within oscillating water column (OWC) systems, which convert the sea-wave motion into a bi-directional flow of air. The Wells turbine transforms the energy of the flow of air into mechanical energy, by means of the aerodynamic forces that are generated on the blades by the relative air motion. A schematic of OWC system and Wells turbine are given in Figure~\ref{fig:owc-wells}. 

As the mass-flow passing though the turbine is alternate and periodic, the blade experiences a continuous variation in incidence angle, i.e. it operates under dynamic conditions. Several authors have discussed the presence of a \emph{hysteretic loop} when representing turbine performance as a function of the flow coefficient: aerodynamic forces acting on the blade were found to be larger during the deceleration phase (when the mass-flow through the turbine is decreasing) than during the acceleration phase (when it is increasing). 

The presence of hysteresis was first discovered in experimental studies conducted on laboratory devices, where a piston moving inside a large cylinder was used to replicate the dynamic operating conditions typical of OWC-installed turbines \cite{Inoue1986, Raghunathan1987, Kaneko1991, Setoguchi1998}, and on full-scale systems~\cite{Thakker2008}. In other studies \cite{Hyun1993, camporeale, puddu_exp, Puddu2014}, the hysteresis appeared negligible. Puddu~\emph{et al.} \cite{puddu_exp, Puddu2014}, in particular, highlighted how the hysteretic effect is significantly reduced when turbine performance is represented as a function of flow parameters measured in the proximity of the rotor.

The (generally accepted) explanation on the origin of the hysteresis was found by means of numerical (CFD) simulations, conducted on a domain consisting of a passage of the annular duct housing the turbine (i.e. chamber and moving piston were not simulated). Kinoue \emph{et al.} \cite{Kinoue2003} attributed the difference in performance between acceleration and deceleration phases to the interaction of trailing edge vortices, shed by the blade due to the variation in flow incidence and opposite in sign during the two phases, with the blade circulation, and another vortical structure that develops near the blade's suction surface. The same explanation is given in other papers by the same group (among many others \cite{Kim2002,Setoguchi2003,Kinoue2004,Kinoue2004a,Mamun2004,Mamun2005,Kinoue2007}), where the phenomenon is compared to the one occurring in rapidly moving airfoils (this subject will be discussed in detail in Section \ref{sec:airfoils}), concluding that the cause has necessarily to be different, given the opposite direction of the hysteretic loops in the two problems, i.e. counter-clockwise in Wells turbines and clockwise in oscillating airfoils (this statement is actually incorrect \cite{Ericsson1988}). The presence of dynamic effects in Wells turbine was the focus of the recent numerical investigations by \cite{Shehata2016, Shehata2017, Shehata2017a, Shehata2017b, Shehata2018, Hu2018}, who gave essentially very similar explanations. In this short note, it will be demonstrated how these results are in contradiction with well established literature on rapidly moving airfoils and wings, and are likely to be caused by errors in the analysis of the experiments and in the setup of the numerical simulations.

\section{Hysteresis in Rapidly Moving Airfoils and Wings}
\label{sec:airfoils}

As highlighted in Section \ref{sec:wells}, the incidence of the flow on a Wells turbine changes continuously during its normal operation. This is similar to what happens in rapidly moving airfoils (pitching or plunging), a phenomenon that has been widely studied since the 1930s \cite{Kramer1932}, given its importance in rotating machinery (wind turbines, compressors and helicopter rotors) \cite{Leishman1990} and animal propulsion (insects, birds, and fish) \cite{Anderson1998}. Significant efforts have been devoted to the study of this problem by NASA in the 1970s and 1980s \cite{Carr1977, McAlister1978, McCroskey1981, McAlister1983, Seto1985}, although several aspects are still at present under investigation 
\cite{Kaufmann2017, Visbal2017, Lee2018, Medina2018, Zeyghami2018, VanBuren2018, Gursul2018}.

A clear explanation of the causes of the hysteresis in rapidly moving airfoils is given by Ericsson and Reding \cite{Ericsson1988}. The problem is governed by three distinct phenomena: 
\begin{enumerate}
\item the interaction of wake vorticity with the airfoil circulation (opposite and concordant in sign during pitch-up and pitch-down, respectively), which determines a lower effective flow incidence and hence a time-lag, responsible for a counter-clockwise loop in aerodynamic performance plots ($c_l-\alpha$, $c_d-\alpha$, and $c_m-\alpha$).
\item the moving wall determines an energization of the boundary layer, that is able to withstand  a larger pressure gradient before separation, with the effect of an increase in stall angle.
\item the generation of a leading edge vortex (LEV), associated with a discontinuous change in circulation (i.e. flow separation) that interacts with the blade suction surface causing, during its passage, a sharp suction peak and a temporary increase in lift. After the passage of the LEV, the airfoil experiences a sudden drop in lift and an increase in drag and pitching moment.
\end{enumerate}

The magnitude of these phenomena depends mainly on the non-dimensional (or reduced) frequency $k$, which is a ratio of the characteristic times of flow passage and airfoil motion \cite{Carr1977, McCroskey1981}. 

\begin{equation}
k=\frac{\pi fc}{U_{\infty}}
\label{eq:f}
\end{equation}
In equation \ref{eq:f}, $f$ is the frequency of oscillation (pitching or plunging), $c$ the airfoil's chord, and $U_{\infty}$ the free stream velocity.
Reynolds number and amplitude of the oscillation are also important, but mainly for determining whether the airfoil exceeds, during its movement, the static stall angle, which in turns determines the formation of the LEV. If the static stall angle is not exceeded, Reynolds number and oscillation amplitude are of secondary importance \cite{Leishman1984}. Mach number effects are also minimal, provided that shock waves are absent \cite{McAlister1978}.

When the airfoil movement does not cause flow separation (i.e. the static stall angle is not exceeded), only the first phenomenon (a) can be present. The vorticity shed by the blade interacts with the blade circulation causing a time-lag in the attainment of the static forces and hence a counter-clockwise hysteretic loop. At reduced frequencies below 0.08, the phase angle produced by shed vorticity is well approximated by the linear relation $\phi=3k$ \cite{Ericsson1988}. It can be easily verified that non-dimensional frequencies well above 10$^{-2}$ are required for this effect to be noticeable.

If the static stall angle is exceeded, there will be both a delay of stall (with respect to static performance) caused by boundary layer improvement (b), and the generation of the LEV, that is convected by the free-stream velocity and therefore interacts with the airfoil for a fraction of the period proportional to $k$. The effect is a clockwise loop in the $c_l-\alpha$ curve, caused by the increase in suction caused by the LEV and by a delayed reattachment of the boundary layer after stall.
These effects start to be important only for $k>4\times 10^{-3}$ \cite{McAlister1978, Seto1985}.

The co-existence of these effects can generate, at high non-dimensional frequencies, the appearance of a bow in the airfoil's aerodynamic curves \cite{McCroskey1981, McAlister1983}.

Wells turbines operate at very low non-dimensional frequencies (lower than 10$^{-3}$ \cite{Setoguchi1998, Shehata2016, Hu2018}), well below the value that is necessary to produce hysteresis in rapidly pitching or plunging airfoils~\cite{McAlister1978, Seto1985, Ericsson1988}, especially if the stall angle is not exceeded, which is what should happen in a well-designed OWC. This fundamental aspect, i.e. the difference in non-dimensional frequency between Wells turbine and oscillating airfoils, was mentioned in several early studies \cite{Setoguchi1998, Setoguchi2003, Kinoue2003}, but seems to have been forgotten in more recent ones \cite{Shehata2016, Hu2018}. The main justification for the apparent inconsistency was found by \cite{Setoguchi1998, Kinoue2003} in the opposite direction of hysteretic loop in Wells turbines (counter-clockwise) and oscillating airfoils (clockwise), that in their opinion can only be caused by a different phenomenon. This justification is undermined by an incorrect analysis of the literature on oscillating airfoils and wings, as it is clear that (in the absence of stall) hysteretic loops are indeed counter-clockwise \cite{Carr1977, McAlister1978}, as it is for Wells turbines. 

\section{Revisiting the Cause of the Hysteresis}

In light of the discrepancies outlined in Section~\ref{sec:airfoils}, the authors of this note have considered important to verify the numerical results presented in older studies, noticing how none of them \cite{Kim2002,Setoguchi2003,Kinoue2004,Kinoue2004a,Mamun2004,Mamun2005,Kinoue2007,Shehata2016, Shehata2017, Shehata2017a, Shehata2017b, Shehata2018, Hu2018} presented an accurate analysis of the numerical errors due to temporal discretization. This appeared curious to say the least, given the importance of excluding any phase errors \cite{JFEpolicy, AIAAJpolicy, JoApolicy} in a study that deals with hysteresis. In fact, numerical phase errors can be confused with real dynamic phenomena, leading to wrong physical conclusions.

By reproducing the same numerical analyses of ~\cite{Kim2002,Setoguchi2003,Kinoue2004,Kinoue2004a,Mamun2004,Mamun2005,Kinoue2007}, i.e. the (isolated) turbine without the connected OWC chamber, the authors of this note verified how the alleged hysteresis was caused only by an incorrect choice of the temporal discretization \cite{Ghisu_JTS, Ghisu_JFE, Ghisu_JoPaE}. 
If numerical simulations are conducted with the appropriate time step, hysteretic effects are negligible, as it is for airfoils oscillating at extremely low non-dimensional frequencies (lower than 10$^{-3}$).

The fact that hysteretic loops appeared in experimental analyses, and not in numerical simulations conducted for an isolated turbine, meant that the cause of the hysteresis had to be found in a different phenomenon. By simulating the full experimental setup (moving piston, chamber, and turbine) Ghisu \emph{et al.} \cite{Ghisu_JFE, Ghisu_JoPaE} showed how the hysteresis was caused by the capacitive behavior of the air mass in the OWC chamber. Only if turbine performance had been represented as a function of local flow parameters, as in the experiments of Puddu \emph{et al.} \cite{Puddu2014}, dynamic effects in the turbine could have been isolated from dynamic effects in the overall OWC, and they would have been found to be negligible \cite{Ghisu_JFE}.

\begin{figure}[!ht]
 \centering
{
\psfrag{p}{\scriptsize{$p_u$}}
\psfrag{m}{\scriptsize{$p_m$}}
\psfrag{tube}{\scriptsize{tube}}
\psfrag{connecting}{\scriptsize{connecting}}
\psfrag{volume}{\scriptsize{volume}}
\psfrag{transducer}{\scriptsize{transducer}}
\psfrag{WellsTurbine}{\scriptsize{Wells turbine}}
\subfigure[]{\includegraphics[width=.25\columnwidth]{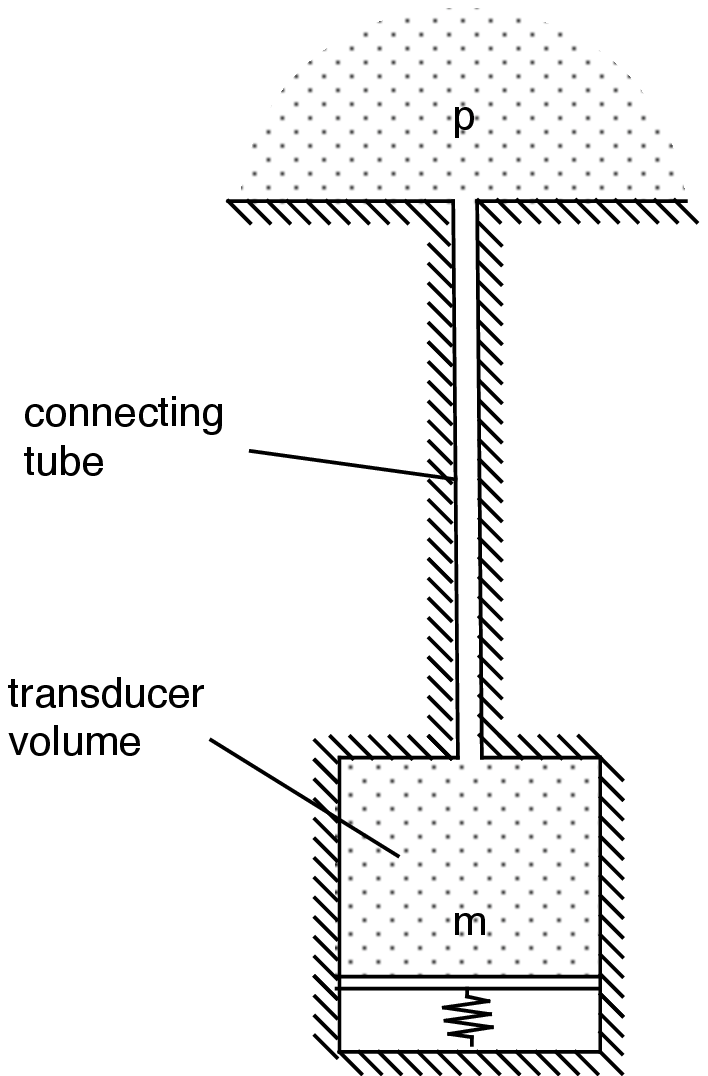}\label{fig:tubetrans}}} \hspace{2cm}
 \subfigure[]{
	 \begin{overpic}[width=.45\columnwidth]{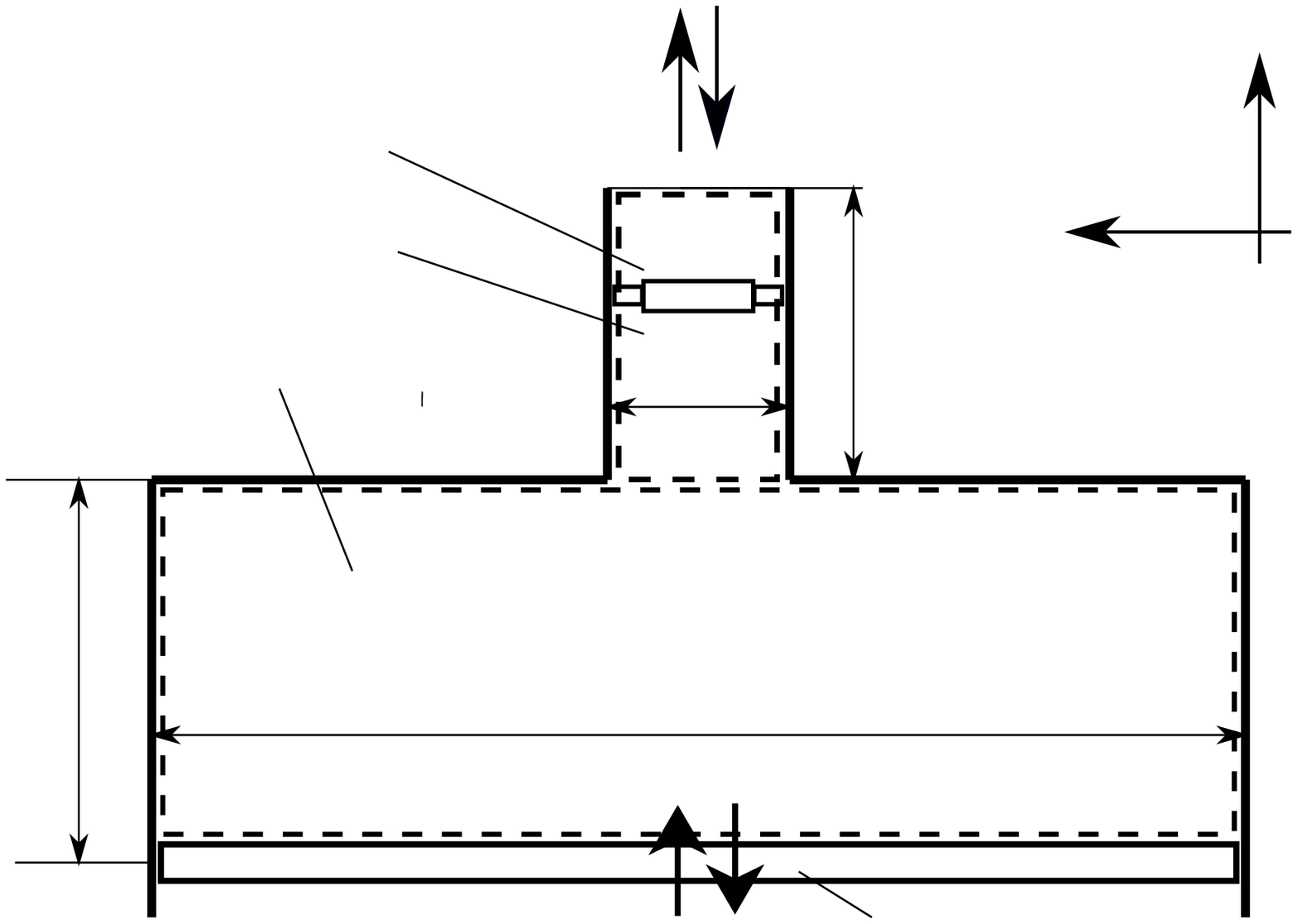}
		\put(-5,51){\scriptsize{\circled{2}}}
		\put(-5,42.7){\scriptsize{\circled{1}}}
		\put(2,51){\scriptsize{Turbine duct}}
		\put(2,42.7){\scriptsize{Air chamber}}
		\put(0,60){\scriptsize{Wells turbine}}
		\put(58,68){\scriptsize{Air}}
		\put(58,64){\scriptsize{mass}}
		\put(58,60){\scriptsize{flow}}
		\put(53,54){\scriptsize{$a$}}
		\put(80,52){\scriptsize{$t$}}
		\put(96,68){\scriptsize{$x$}}
		\put(79,47){\scriptsize{tangential}}
		\put(101,54){\mbox{\rotatebox{90}{\scriptsize{axial}}}}
		\put(51.5,41){\scriptsize{$A_2$}}
		\put(51.5,16){\scriptsize{$A_1$}}
		\put(-5,17){\scriptsize{$h_1(t)$}}
		\put(60,-3){\scriptsize{moving piston}}
	\end{overpic}\label{fig:cv}}
\caption{Schematics of tube-transducer system (left) and laboratory OWC system (right)}
\end{figure}

An interesting and remarkably simple explanation of the real cause of the hysteresis can be given through a comparison with a different problem, the analysis of the dynamic behavior of pressure measurement systems. In fact, pressure transducers are composed of a (variable size) volume connected to the measurement point through a pneumatic line (a narrower tube). A schematic of a tube-transducer arrangement is presented in Figure~\ref{fig:tubetrans}, next to a schematic of the experimental setup used for the evaluation of dynamic effects in Wells turbines (Figure~\ref{fig:cv})~\cite{Setoguchi1998, camporeale, puddu_exp}. The similarity between the two setups is evident.

\tiny
\normalsize

When measuring a dynamically changing pressure with the instrument in Figure \ref{fig:tubetrans}, attention needs to be payed to the delay that can exist between flow conditions just outside the duct (the pressure one seeks to know, $p_u$) and inside the volume (the pressure that is actually measured by the transducer, $p_m$). This is explained in detail in \cite[Chapter 6.6: Dynamic Effects of Volumes and Connecting Tubing]{Doebelin}, where a lumped parameter model (LPM) approach is used to evaluate the phase delay between actual and measured pressures. An adaptation of this approach will be presented next.

An LPM model of the OWC-Wells turbine system can be derived by applying the laws of conservation of mass and axial momentum to the variable air volume in the chamber (\circled{1} in Figure~\ref{fig:cv}) and to the turbine duct (\circled{2}), respectively. 

\begin{equation}
\begin{dcases}
\frac{\dft M_1}{\dft t} = h_1 A_1 \frac{\dft \rho_1}{\dft t} + \rho_1 A_1 \frac{\dft h_1}{\dft t} =
                         -\rho_a V_2 A_{2}\\
\frac{\dft (\rho_2 V_2 A_2 L)}{\dft t} =
                        (p_1 - p_a) A_{2} + F_{x}
\end{dcases}
\label{eqdiff}
\end{equation}

The rate of variation of air mass in the chamber is equal to the mass-flow leaving the control volume through the opening, while the rate of momentum in the turbine duct is equal to the forces acting on the corresponding control volume, i.e. the sum of pressure forces on the boundaries and aerodynamic forces from the turbine ($F_x$). Compressibility of air within the turbine duct (but not in the overall system) has been neglected ($\rho_f=\rho_a=\rho_2$), with the result that the mass-flow through any section of the turbine duct is assumed constant. Forces due to friction on the duct walls have been considered negligible with respect to aerodynamic forces acting on the turbine. 

Wells turbine performance is represented in terms of non-dimensional coefficients of pressure drop $P^*$ and torque $T^*$, as a function of flow coefficient $\phi$, which in the experiment of Setoguchi~\cite{Setoguchi1998} appear to be calculated based on piston speed.
\begin{equation}\label{definizione_TP}
		  P^*    = \frac{\Delta P}{\rho \omega^2 {r_{m}}^2}; \quad
T^*    = \frac {T} {\rho \omega^2 {r_{m}}^5}          ; \quad  
		  \phi_p = \frac{V_{p}}{\omega r_{m}}\frac{A_1}{A_2}; \quad
\end{equation}

The equations in (\ref{eqdiff}) can be written in terms of the non-dimensional coefficients in (\ref{definizione_TP}), by dividing the mass conservation equation by ($\rho_a \omega r_m A_2$) and the momentum equation by ($\rho_a (\omega r_m)^2 A_2$):

\begin{equation}
\begin{dcases}
	\frac{h_1}{r_m} \frac{A_1}{A_2} \frac{\dft (\rho_1 / \rho_a)}{\dft (t\omega)}+ \frac{\rho_1}{\rho_a}\frac{A_1}{A_2}\frac{\dft (h_1/r_m)}{\dft (t\omega)} 
=-\frac{V_2}{\omega r_m} 
\\
\frac{L}{r_m}\frac{ \dft (V_2/(\omega r_m))}{\dft (t \omega)} = \frac{(p_1 - p_{a})}{\rho_a (\omega r_m)^2} + \frac{F_{x}}{\rho_a(\omega r_m)^2 A_2}
\label{eq:ndsystem}
\end{dcases}
\end{equation}

Introducing the following non-dimensional parameters:
\begin{align*}
	&\frac{A_1}{A_2}\frac{\dft(h_1/r_m)}{\dft(t\omega)} =-\phi_p &\quad
	&\frac{V_2}{\omega r_m} =\phi_l &\quad 
	&\frac{p_1-p_a}{\rho_a(\omega r_m)^2} =P^* 
	\\
	&\frac{\rho_1}{\rho_a} = \rho^* = \frac{\gamma(\omega r_m)^2}{a^2} P^* +1 &\quad 
	&\frac{F_x}{\rho_a(\omega r_m)^2 A_2} = c_x &\quad
	&t\,\Omega = \frac{t \omega}{\omega/\Omega}= t^* 
\end{align*}

equation~(\ref{eq:ndsystem}) becomes: 
\begin{equation}
\begin{dcases}
\frac{h_1 A_1\Omega}{A_2}\frac{\gamma(\omega r_m)}{a^2} \frac{\dft P^*}{\dft t^*}
	- \left(\frac{\gamma(\omega r_m)^2}{a^2}P^*+1\right)\phi_p
=-\phi_l
\\
\frac{L}{r_m}\frac{\Omega}{\omega}\frac{\dft \phi_l}{\dft t^*} = P^*+c_x
\label{eq:ndsystem-s}
\end{dcases}
\end{equation}
where $\phi_l$ is a local flow coefficient calculated based on the axial velocity of the flow in the turbine duct, and $c_x$ a non-dimensional coefficient for the aerodynamic axial turbine force. Equation (\ref{eq:ndsystem-s}) represents a system of two first order non-linear ordinary differential equations with $P^*$ and $\phi_l$ representing the unknowns and $\phi_p$ the external forcing. $c_x$  needs to be provided as a function of other working parameters. 

Assuming turbine aerodynamic hysteresis to be negligible, and considering that the force coefficient $c_x$ can be well approximated as a linear function of $\phi_l$ ($c_x=c_{x,\phi}\phi$), while $T^*$ can be approximated as a quadratic function of $\phi_l$ \cite{Ghisu_JFE},
equation (\ref{eq:ndsystem-s}) can be converted to a single second order differential equation:

\begin{equation}
		  \frac{L}{r_m}\frac{\Omega}{\omega}\frac{\dft ^2\phi_l}{\dft t^{*2}} 
		  + c_{x,\phi}\frac{\dft \phi_l}{\dft t^*} 
		  + \frac{a^2}{\gamma(\omega r_m)(h_{1} \Omega)}\frac{A_2}{A_1}\phi_l
		  = \frac{a^2}{\gamma(\omega r_m)(h_{1} \Omega)}\frac{A_2}{A_1}\left(1+\frac{\gamma(\omega r_m)^2}{a^2}P^*\right)\phi_p
\label{eq:2nd_ord1}
\end{equation}
In order to be solved analytically, equation~(\ref{eq:2nd_ord1}) needs to be linearized. This can be done by assuming:

\begin{align}
	&h_{1}\approx h_{10} \text{ (its value at rest)} 
	&\left(1+\frac{\gamma(\omega r_m)^2}{a^2}P^*\right) \approx 1
\end{align}

After linearization, equation (\ref{eq:2nd_ord1}) becomes:

\begin{equation}
		  \underbrace{\frac{L}{r_m}\frac{\Omega}{\omega}}_{A}\frac{\dft ^2\phi_l}{\dft t^{*2}} 
		  + \underbrace{c_{x,\phi}}_{B}\frac{\dft \phi_l}{\dft t^*} 
		  + \underbrace{\frac{a^2}{\gamma(\omega r_m)(h_{10} \Omega)}\frac{A_2}{A_1}}_{C}\phi_l
		  = \underbrace{\frac{a^2}{\gamma(\omega r_m)(h_{10} \Omega)}\frac{A_2}{A_1}}_{D} \phi_p
\label{eq:2nd_ord2}
\end{equation}

The solution to equation (\ref{eq:2nd_ord2}) can be seen in terms of its transfer function $G(\Omega / \Omega_n)$:
\begin{equation}
	G\left(\frac{\Omega}{\Omega_n}\right) = \frac{\phi_l}{\phi_p}= \frac{D}{-A+Bj+C}=
	\frac{\frac{D}{C}}{-\frac{A}{C}+\frac{B}{C}j+1}=
	\frac{\frac{D}{C}}{-\left(\frac{\Omega}{\Omega_n}\right)^2 +1+2\zeta\left(\frac{\Omega}{\Omega_n}\right)j}
	\label{eq:secondorder_s}
\end{equation}
In the above equations, $\Omega_n$ is the angular natural frequency and $\zeta$ the damping ratio of the system:
\begin{align}
	&\Omega_n=\sqrt{\frac{C}{A}}\Omega=a\sqrt{\frac{1}{\gamma h_{10}L}\frac{A_2}{A_1}} 
	&2\zeta=\frac B C \frac{\Omega_n}{\Omega}=\frac B C \sqrt{\frac C A} = \frac B {\sqrt{AC}}=\frac {c_{x,\phi}}{\sqrt{\frac{L}{h_{10}}\frac{A_2}{A_1}\frac{1}{\gamma}}\frac{a}{\omega r_m}}
\end{align}

The solution to equation (\ref{eq:2nd_ord2}) is therefore:
\begin{equation}
	\phi_l=\phi_{l0}\,e^{jt^*+\xi}
\end{equation}

where:
\begin{align}
	&\phi_{l0}=|\phi_l|=|\phi_p|\left|G\left(\frac{\Omega}{\Omega_n}\right)\right| \\
	&\phi_{p}=\phi_{p0}\,e^{jt^*} \\
	&\left|G\left(\frac{\Omega}{\Omega_n}\right)\right|= \frac{\frac{D}{C}}{\sqrt{\left[\left(-\frac{\Omega}{\Omega_n}\right)^2+1\right]^2+\left[2\zeta\left(\frac{\Omega}{\Omega_n}\right)\right]^2}}=\frac{D}{\sqrt{\left(C-A\right)^2+B^2}} \label{eq:gs1}\\
	&\xi = \tan^{-1}\left(\frac{-2\zeta\frac{\Omega}{\Omega_n}}{-\left(\frac{\Omega}{\Omega_n}\right)^2+1}\right)=\tan^{-1}\left(\frac{-B}{C-A}\right)= 
 \tan^{-1}\left(\frac{c_{x,\phi}}{\frac{L}{r_m}\frac{\Omega}{\omega}-\frac{a^2}{\gamma(\omega r_m)(h_{10} \Omega)}\frac{A_2}{A_1}}\right)   \label{eq:xi1}
\end{align}

Equation (\ref{eq:xi1}) deserves some attention. Because of the damping produced by the first order term in equation~(\ref{eq:2nd_ord2}) (the resistance produced by the turbine), a delay exists between piston movement and mass-flow passing in the turbine duct. It will be shown how this \emph{OWC hysteresis} is by far the largest contribution to the hysteresis measured in the experiments of \cite{Inoue1986, Raghunathan1987, Kaneko1991, Setoguchi1998, Thakker2008}.

Table \ref{tab:hyst} presents a numerical comparison between the OWC hysteresis and the aerodynamic hysteresis of the turbine. The former has been estimated from equation (\ref{eq:xi1}), the latter from the equations given by \cite{Ericsson1988}. The comparison has been made for three experiments, characterized by different turbine solidities.

\begin{table}[!h]
\centering
\caption{Geometrical and operating data for Setoguchi's experiments \cite{Setoguchi1998}}\label{tab:hyst}
\begin{tabular}{lccc}
\toprule
	Experiment & 1 & 2 & 3 \\
			chamber diameter [m] &  \multicolumn{3}{c}{1.4 m} \\
			rotor tip diameter [mm] &  \multicolumn{3}{c}{300 mm} \\
			rotor hub diameter [mm] &  \multicolumn{3}{c}{210 mm}\\
			tip clearance [mm] &  \multicolumn{3}{c}{1 mm}\\
			chord length $c$ [mm] &  \multicolumn{3}{c}{90 mm}\\
			sweep ratio [-] &  \multicolumn{3}{c}{0.417} \\
			number of blades [-]&  5 & 6 & 7 \\
			blade profile [-] &  \multicolumn{3}{c}{NACA0020} \\
			solidity at tip radius $\sigma$ [-]&  0.48 & 0.57 & 0.67\\
			$c_{x,\phi}[-]$ & 2.05 & 3.11& 5.67 \\  
			rotational speed [rpm]&  \multicolumn{3}{c}{2500 rpm}\\
			piston frequency $f$ [s$^{-1}$]&  \multicolumn{3}{c}{$6$ s}\\
			Reynolds number $Re$ [-]&  \multicolumn{3}{c}{$2 \times 10^5$}\\
			Mach number $M$ [-]&  \multicolumn{3}{c}{0.1}\\
			turbine non-dimensional frequency $k$ [-]&  \multicolumn{3}{c}{0.0012}\\
			phase delay due to turbine [-] & \multicolumn{3}{c}{0.0036}\\
			phase delay due to OWC [-]& 0.036 & 0.054 & 0.098 \\
\bottomrule
\end{tabular}
\end{table}

From the above results, it is clear the contribution to phase delay, or hysteresis, between piston speed and turbine force coefficients, given by turbine aerodynamics is at least one order of magnitude lower than the the contribution given by the OWC's capacitive behavior. In other words, it is absolutely paramount that the capacitive behavior of the OWC is taken into consideration in experiments, and this does not appear to have been done in \cite{Inoue1986, Raghunathan1987, Kaneko1991, Setoguchi1998}.

\begin{figure}
\centering
	\subfigure[$\sigma=0.48$]{
                  {\psfrag{X}[l]{\tiny{$\phi_p$}}
\psfrag{Y}{\tiny{$P^*$}}
\psfrag{EXP}[][l]{\tiny{EXP}}
\psfrag{CFD}[][l]{\tiny{CFD}}
\psfrag{LPM}[][l]{\tiny{LPM}}
	\includegraphics[width=0.35\textwidth]{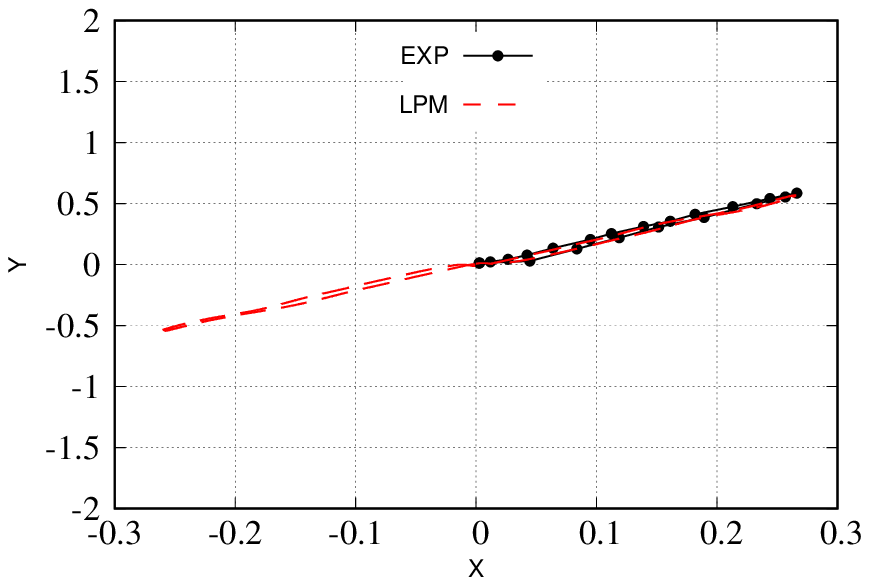}}
                  {\psfrag{X}[l]{\tiny{$\phi_p$}}
\psfrag{Y}{\tiny{$T^*$}}
\psfrag{EXP}[][l]{\tiny{EXP}}
\psfrag{CFD}[][l]{\tiny{CFD}}
\psfrag{LPM}[][l]{\tiny{LPM}}
	\includegraphics[width=0.35\textwidth]{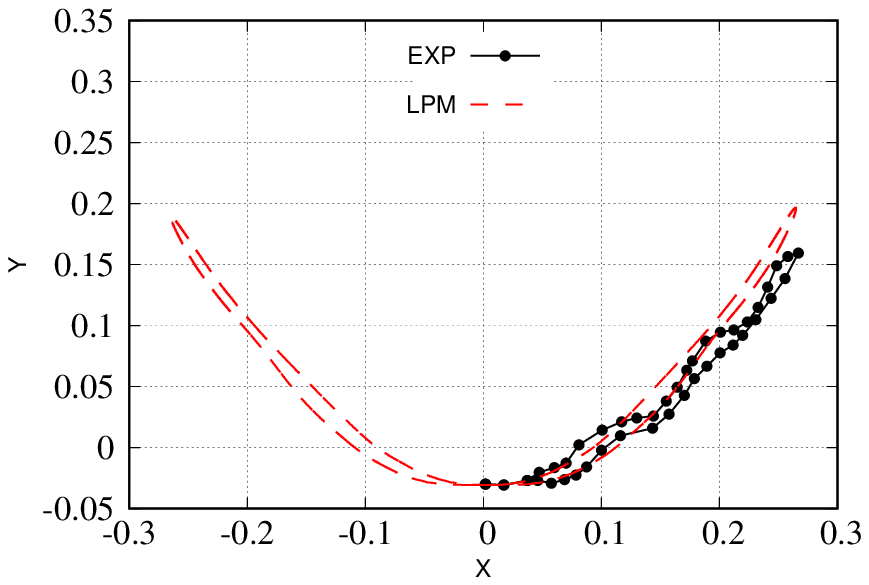}}}
	\subfigure[$\sigma=0.57$]{
                  {\psfrag{X}[l]{\tiny{$\phi_p$}}
\psfrag{Y}{\tiny{$P^*$}}
\psfrag{EXP}[][l]{\tiny{EXP}}
\psfrag{CFD}[][l]{\tiny{CFD}}
\psfrag{LPM}[][l]{\tiny{LPM}}
	\includegraphics[width=0.35\textwidth]{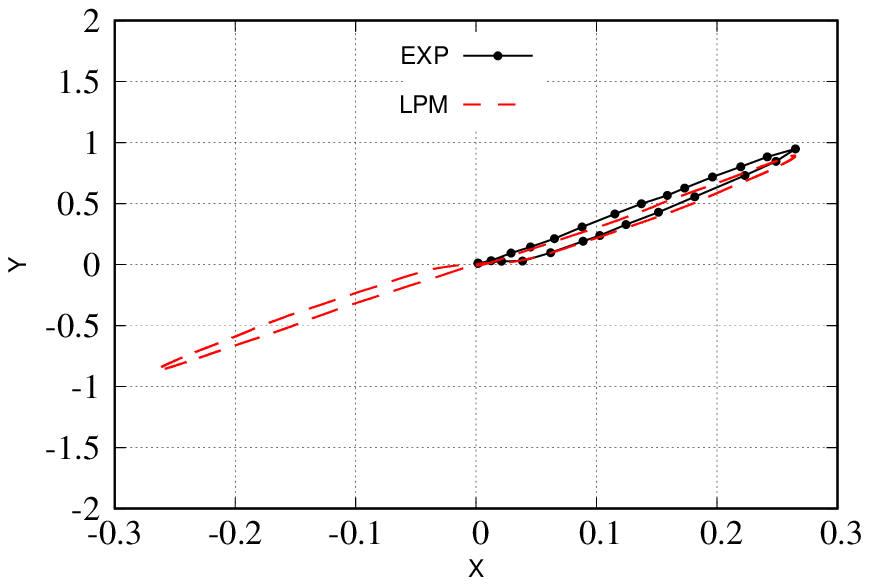}}
                  {\psfrag{X}[l]{\tiny{$\phi_p$}}
\psfrag{Y}{\tiny{$T^*$}}
\psfrag{EXP}[][l]{\tiny{EXP}}
\psfrag{CFD}[][l]{\tiny{CFD}}
\psfrag{LPM}[][l]{\tiny{LPM}}
	\includegraphics[width=0.35\textwidth]{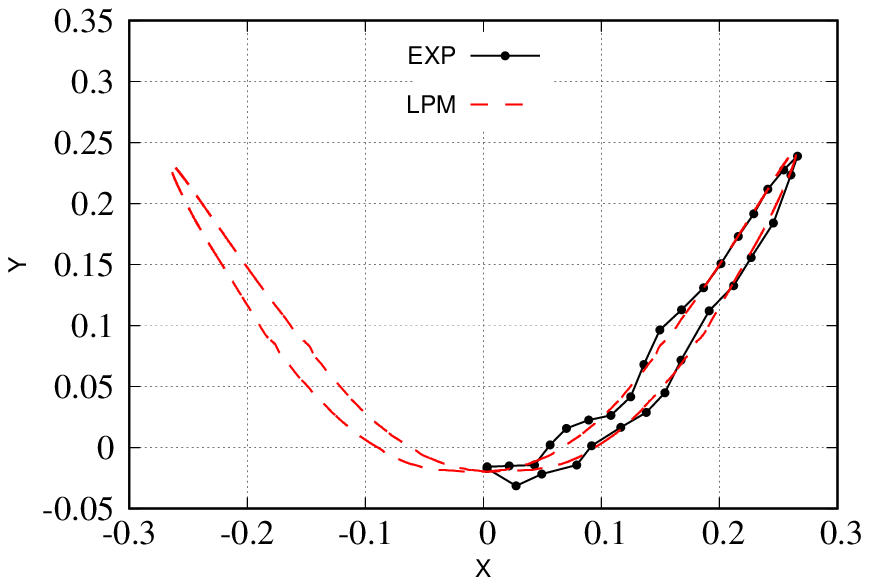}}}
	\subfigure[$\sigma=0.67$]{
                  {\psfrag{X}[l]{\tiny{$\phi_p$}}
\psfrag{Y}{\tiny{$P^*$}}
\psfrag{EXP}[][l]{\tiny{EXP}}
\psfrag{CFD}[][l]{\tiny{CFD}}
\psfrag{LPM}[][l]{\tiny{LPM}}
	\includegraphics[width=0.35\textwidth]{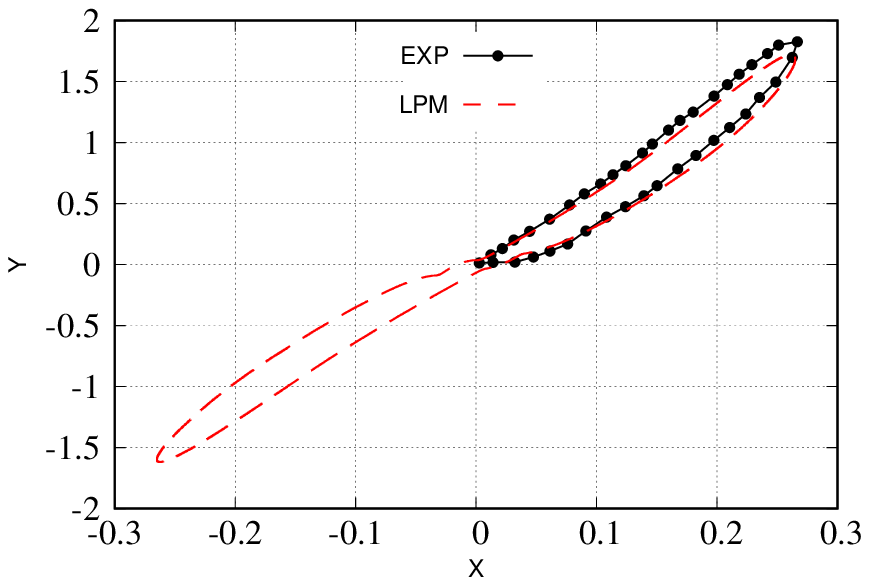}}
                  {\psfrag{X}[l]{\tiny{$\phi_p$}}
\psfrag{Y}{\tiny{$T^*$}}
\psfrag{EXP}[][l]{\tiny{EXP}}
\psfrag{CFD}[][l]{\tiny{CFD}}
\psfrag{LPM}[][l]{\tiny{LPM}}
	\includegraphics[width=0.35\textwidth]{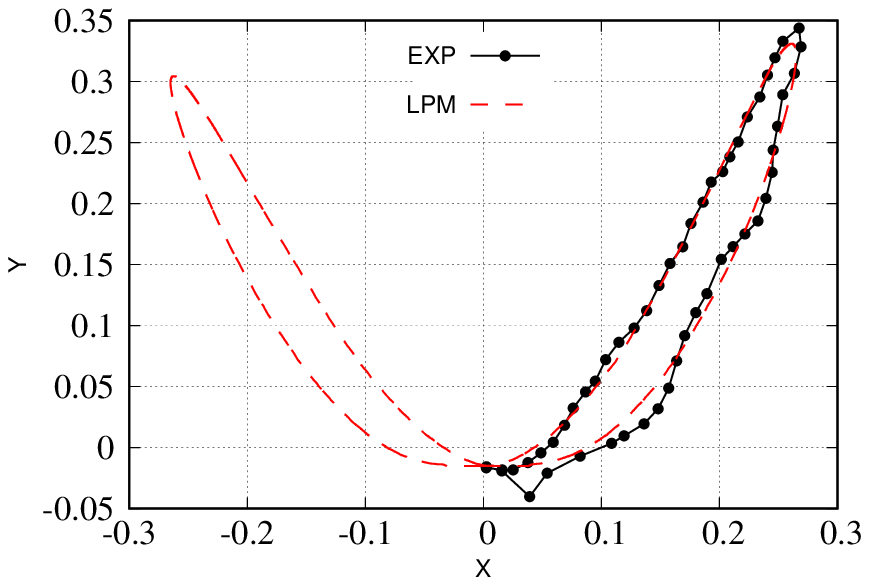}}}
\caption{Comparison of LPM results with experimental data from \cite{Setoguchi1998}, for different turbine solidities}
\label{fig:PTphi}
\end{figure}

Figure~\ref{fig:PTphi} compares the results obtained with a numerical solution of equation (\ref{eq:2nd_ord2}), using a semi-implicit time-marching scheme with a time-step of 10$^{-4}$ s, with the experimental results of \cite{Setoguchi1998}. The time-step has been selected carefully to avoid the presence of phase errors due to temporal discretization. The LPM approach (without any account for aerodynamic turbine hysteresis) is able to predict with remarkable accuracy the hysteretic loop found in the experiments. The turbine solidity has a direct effect on the slope of the $c_x$ vs. $\phi$ curve, i.e. on the damping term in equation (\ref{eq:2nd_ord2}).

This is a clear demonstration that the hysteresis is caused solely by capacitive effects in the OWC and not by a turbine aerodynamic hysteresis, which is negligible at the non-dimensional frequency Wells turbine operate at. It is therefore important, when analyzing Wells turbines installed in OWC systems, to consider that dynamic effects can arise from different sources, which need to be singled out before examining the unlikely presence of any aerodynamic hysteresis in the turbine.

\FloatBarrier
\newpage
\section{Conclusions}

The presence of aerodynamic hysteresis in Wells turbines has been the subject of a large number of publications in the last decades. Its presence had been discovered in experimental analyses, which studied the turbine behavior in laboratory setups that used a mechanic piston inside a lareg cylinder to reproduce the periodic mass-flow through the rotor. The commonly accepted explanation was found using CFD simulations, that reproduced only part of the laboratory experiment, i.e. the turbine rotor, neglecting the importance of the large chamber used to reproduce the periodic operating conditions. 

This note suggests a re-examination of the origin of the alleged hysteresis, based on two anomalies. First,  during its operation, a Wells turbine experiences a continuous change in incidence angle, in a way that is not dissimilar to what happens to oscillating airfoils and wings, but at non-dimensional frequencies too low to generate any visible dynamic effects (i.e. hysteresis) in oscillating airfoils.
Second, none of the many articles that studied this phenomenon numerically conducted an analysis of temporal discretization errors, fundamental to ensure the absence of spurious phase errors.

Recently, the authors of this note re-examined the problem using a CFD analysis, and conducted a study of the sensitivity of the results to the temporal discretization. They highlighted how the hysteresis reported by other authors is likely to be caused by numerical errors. The result of a negligible aerodynamic hysteresis in Wells turbines would be in accordance with the large existing literature on oscillating airfoils and wings. 

In this work, it is shown that the real cause of the hysteresis can be found in a different phenomenon, neglected in previous experimental and numerical analyses: the phase difference between piston movement and mass-flow in the turbine duct, caused the capacitive behavior of the large OWC chamber. A lumped parameter model of the system is therefore used to estimate this phase delay, and the results are compared to the experimental analysis that first reported the hysteresis under consideration.  It is shown that the OWC system hysteresis completely explains the experimental results and that the aerodynamic hysteresis is likely to be at least one order of magnitude lower than the OWC hysteresis, hence negligible. The errors produced by previous analyses could have been avoided with an attentive analysis of the experimental setup and with a proper verification of numerical results. 

\section*{Acknowledgements}

This work has been funded by the Regione Autonoma Sardegna under grant F72F16002880002 (L.R. 7/2007 n. 7 - year 2015).


\bibliography{mybib}  

\begin{thebibliography}{47}
\newcommand{\enquote}[1]{``#1''}
\providecommand{\natexlab}[1]{#1}
\providecommand{\url}[1]{\texttt{#1}}
\providecommand{\urlprefix}{URL }
\expandafter\ifx\csname urlstyle\endcsname\relax
  \providecommand{\doi}[1]{doi:\discretionary{}{}{}#1}\else
  \providecommand{\doi}{doi:\discretionary{}{}{}\begingroup
  \urlstyle{rm}\Url}\fi

\bibitem[{Inoue et~al.(1986)Inoue, Kaneko, Setoguchi, and
  Shimamoto}]{Inoue1986}
Inoue, M., Kaneko, K., Setoguchi, T., and Shimamoto, K., \enquote{Studies on
  {W}ells Turbine for Wave Power Generator(Part 4: Starting and Running
  Characteristics in Periodically Oscillating Flow),} \emph{Bulletin of JSME},
  Vol.~29, No. 250, 1986, pp. 1177--82.
\newblock \doi{10.1248/cpb.37.3229}.

\bibitem[{Raghunathan et~al.(1987)Raghunathan, Setoguchi, and
  Kaneko}]{Raghunathan1987}
Raghunathan, S., Setoguchi, T., and Kaneko, K., \enquote{Hysteresis on {W}ells
  turbine blades,} \emph{ASME Fluids Engineering Conference}, Cincinnati, USA,
  1987.

\bibitem[{Kaneko et~al.(1991)Kaneko, Setoguchi, Hamakawa, and
  Inoue}]{Kaneko1991}
Kaneko, K., Setoguchi, T., Hamakawa, H., and Inoue, M., \enquote{{Biplane axial
  turbine for wave power generator},} \emph{International Journal of Offshore
  and Polar Engineering}, Vol.~1, No.~2, 1991, pp. 122--128.

\bibitem[{Setoguchi et~al.(1998)Setoguchi, Takao, and Kaneko}]{Setoguchi1998}
Setoguchi, T., Takao, M., and Kaneko, K., \enquote{{Hysteresis on Wells turbine
  characteristics in reciprocating flow},} \emph{International Journal of
  Rotating Machinery}, Vol.~4, No.~1, 1998, pp. 17--24.

\bibitem[{Thakker and Abdulhadi(2008)}]{Thakker2008}
Thakker, A., and Abdulhadi, R., \enquote{{The performance of Wells turbine
  under bi-directional airflow},} \emph{Renewable Energy}, Vol.~33, No.~11,
  2008, pp. 2467--2474.

\bibitem[{Hyun et~al.(1993)Hyun, Suh, and Lee}]{Hyun1993}
Hyun, B.~S., Suh, J.~S., and Lee, P.~M., \enquote{{Investigation on the
  aerodynamic performance of a Wells turbine for ocean wave energy
  absorption},} \emph{Transactions of the Society of Naval Architects of
  Korea}, 1993.

\bibitem[{Camporeale and Filianoti(2009)}]{camporeale}
Camporeale, S., and Filianoti, P., \enquote{Behaviour of a Small {W}ells
  Turbine under Randomly Varying Oscillating Flow,} \emph{Proceedings of the
  8th European Wave and Tidal Energy Conference}, Uppsala, Sweden, 2009, pp.
  690--696.

\bibitem[{Paderi and Puddu.(2013)}]{puddu_exp}
Paderi, M., and Puddu., P., \enquote{Experimental Investigation in a {W}ells
  Turbine Under Bi-directional Flow,} \emph{Renewable Energy}, Vol.~57, 2013,
  pp. 570--576.
\newblock \doi{10.1016/j.renene.2013.02.016}.

\bibitem[{Puddu et~al.(2014)Puddu, Paderi, and Manca}]{Puddu2014}
Puddu, P., Paderi, M., and Manca, C., \enquote{Aerodynamic Characterization of
  a {W}ells Turbine under Bi-directional Airflow,} \emph{Energy Procedia},
  Vol.~45, 2014, pp. 278--287.
\newblock \doi{10.1016/j.egypro.2014.01.030}.

\bibitem[{Kinoue et~al.(2003)Kinoue, Setoguchi, Kim, Kaneko, and
  Inoue}]{Kinoue2003}
Kinoue, Y., Setoguchi, T., Kim, T.~H., Kaneko, K., and Inoue, M.,
  \enquote{Mechanism of Hysteretic Characteristics of {W}ells Turbine for Wave
  Power Conversion,} \emph{Journal of Fluids Engineering}, Vol. 125, No.~2,
  2003, pp. 302--307.
\newblock \doi{10.1115/1.1538629}.

\bibitem[{Kim et~al.(2002)Kim, Setoguchi, Takao, Kaneko, and
  Santhakumar}]{Kim2002}
Kim, T.~H., Setoguchi, T., Takao, M., Kaneko, K., and Santhakumar, S.,
  \enquote{{Study of turbine with self-pitch-controlled blades for wave energy
  conversion},} \emph{International Journal of Thermal Sciences}, Vol.~41,
  No.~1, 2002, pp. 101--107.
\newblock \doi{10.1016/S1290-0729(01)01308-4}.

\bibitem[{Setoguchi et~al.(2003)Setoguchi, Kinoue, Kim, Kaneko, and
  Inoue}]{Setoguchi2003}
Setoguchi, T., Kinoue, Y., Kim, T., Kaneko, K., and Inoue, M.,
  \enquote{Hysteretic characteristics of Wells turbine for wave power
  conversion,} \emph{Renewable Energy}, Vol.~28, No.~13, 2003, pp. 2113--2127.
\newblock \doi{10.1016/S0960-1481(03)00079-X}.

\bibitem[{Kinoue et~al.(2004{\natexlab{a}})Kinoue, Kim, Setoguchi, Mohammad,
  Kaneko, and Inoue}]{Kinoue2004}
Kinoue, Y., Kim, T.~H., Setoguchi, T., Mohammad, M., Kaneko, K., and Inoue, M.,
  \enquote{{Hysteretic characteristics of monoplane and biplane Wells turbine
  for wave power conversion},} \emph{Energy Conversion and Management},
  Vol.~45, No. 9-10, 2004{\natexlab{a}}, pp. 1617--1629.

\bibitem[{Kinoue et~al.(2004{\natexlab{b}})Kinoue, Setoguchi, Kim, Mamun,
  Kaneko, and Inoue}]{Kinoue2004a}
Kinoue, Y., Setoguchi, T., Kim, T., Mamun, M., Kaneko, K., and Inoue, M.,
  \enquote{Hysteretic characteristics of the {W}ells turbine in a deep stall
  condition,} \emph{Proceedings of the Institution of Mechanical Engineers Part
  M: Journal of Engineering for the Maritime Environment}, Vol. 218, No.~3,
  2004{\natexlab{b}}, pp. 167--173.
\newblock \doi{10.1243/1475090041737967}.

\bibitem[{Mamun et~al.(2004)Mamun, Kinoue, Setoguchi, Kim, Kaneko, and
  Inoue}]{Mamun2004}
Mamun, M., Kinoue, Y., Setoguchi, T., Kim, T., Kaneko, K., and Inoue, M.,
  \enquote{Hysteretic flow characteristics of biplane {W}ells turbine,}
  \emph{Ocean Engineering}, Vol.~31, No. 11-12, 2004, pp. 1423--1435.
\newblock \doi{10.1016/j.oceaneng.2004.03.002}.

\bibitem[{Mamun et~al.(2005)Mamun, Setoguchi, Kinoue, and Kaneko}]{Mamun2005}
Mamun, M., Setoguchi, T., Kinoue, Y., and Kaneko, K., \enquote{Visualization of
  unsteady flow phenomena of {W}ells turbine during hysteresis study,}
  \emph{Journal of Flow Visualization and Image Processing}, Vol.~12, No.~2,
  2005, pp. 111--123.
\newblock \doi{10.1615/JFlowVisImageProc.v12.i2.20}.

\bibitem[{Kinoue et~al.(2007)Kinoue, Mamun, Setoguchi, and Kaneko}]{Kinoue2007}
Kinoue, Y., Mamun, M., Setoguchi, T., and Kaneko, K., \enquote{Hysteretic
  characteristics of {W}ells turbine for wave power conversion (effects of
  solidity and setting angle),} \emph{International Journal of Sustainable
  Energy}, Vol.~26, No.~1, 2007, pp. 51--60.
\newblock \doi{10.1080/14786450701359117}.

\bibitem[{Ericsson and Reding(1988)}]{Ericsson1988}
Ericsson, L.~E., and Reding, J.~P., \enquote{Fluid Mechanics of Dynamic Stall:
  Part 1 Unsteady Flow Concepts,} \emph{Journal of Fluids and Structures},
  Vol.~2, 1988, pp. 1--33.
\newblock \doi{10.1016/S0889-9746(88)80015-X}.

\bibitem[{Shehata et~al.(2016)Shehata, Saqr, Xiao, Shehadeh, and
  Day}]{Shehata2016}
Shehata, A.~S., Saqr, K.~M., Xiao, Q., Shehadeh, M.~F., and Day, A.,
  \enquote{Performance Analysis of {W}ells Turbine Blades Using the Entropy
  Generation Minimization Method,} \emph{Renewable Energy}, Vol.~86, 2016, pp.
  1123--1133.
\newblock \doi{10.1016/j.renene.2015.09.045}.

\bibitem[{Shehata et~al.(2017{\natexlab{a}})Shehata, Xiao, El-Shaib, Sharara,
  and Alexander}]{Shehata2017}
Shehata, A.~S., Xiao, Q., El-Shaib, M., Sharara, A., and Alexander, D.,
  \enquote{{Comparative analysis of different wave turbine designs based on
  conditions relevant to northern coast of Egypt},} \emph{Energy}, Vol. 120,
  2017{\natexlab{a}}, pp. 450--467.
\newblock \doi{10.1016/j.energy.2016.11.091}.

\bibitem[{Shehata et~al.(2017{\natexlab{b}})Shehata, Xiao, Saqr, and
  Naguib}]{Shehata2017a}
Shehata, A.~S., Xiao, Q., Saqr, K.~M., and Naguib, D., A.and~Alexander,
  \enquote{{Passive flow control for aerodynamic performance enhancement of
  airfoil with its application in Wells turbine -- Under oscillating flow
  condition},} \emph{Ocean Engineering}, Vol. 136, 2017{\natexlab{b}}, pp.
  31--53.
\newblock \doi{10.1016/j.oceaneng.2017.03.010}.

\bibitem[{Shehata et~al.(2017{\natexlab{c}})Shehata, Xiao, Selim, Elbatran, and
  Alexander}]{Shehata2017b}
Shehata, A.~S., Xiao, Q., Selim, M.~M., Elbatran, A.~H., and Alexander, D.,
  \enquote{{Enhancement of performance of wave turbine during stall using
  passive flow control: First and second law analysis},} \emph{Renewable
  Energy}, Vol. 113, 2017{\natexlab{c}}, pp. 369--392.
\newblock \doi{10.1016/j.renene.2017.06.008}.

\bibitem[{Shehata et~al.(2018)Shehata, Xiao, Kotb, Selim, Elbatran, and
  Alexander}]{Shehata2018}
Shehata, A., Xiao, Q., Kotb, M., Selim, M., Elbatran, A., and Alexander, D.,
  \enquote{Effect of passive flow control on the aerodynamic performance,
  entropy generation and aeroacoustic noise of axial turbines for wave energy
  extractor,} \emph{Ocean Engineering}, Vol. 157, 2018, pp. 262--300.
\newblock \doi{10.1016/j.oceaneng.2018.03.053}.

\bibitem[{Hu and Li(2018)}]{Hu2018}
Hu, Q., and Li, Y., \enquote{Unsteady {RANS} Simulations of Wells Turbine Under
  Transient Flow Conditions,} \emph{ASME Journal of Offshore Mechanics and
  Arctic Engineering}, Vol. 140(1), 2018.
\newblock \doi{http://dx.doi.org/10.1115/1.4037696}.

\bibitem[{Kramer(1932)}]{Kramer1932}
Kramer, M., \enquote{{Increase in the maximum lift of an airfoil due to a
  sudden increase in its effective angle of attack resulting from a gust},}
  Tech. Rep. NASA Technical Memorandum 678, NASA, 1932.

\bibitem[{Leishman(1990)}]{Leishman1990}
Leishman, J.~G., \enquote{{Dynamic stall experiments on the NACA 23012
  aerofoil},} \emph{Experiments in Fluids}, Vol.~9, No. 1-2, 1990, pp. 49--58.
\newblock \doi{10.1007/BF00575335}.

\bibitem[{Anderson et~al.(1998)Anderson, Streitlien, Barrett, and
  Triantafyllou}]{Anderson1998}
Anderson, J.~M., Streitlien, K., Barrett, D.~S., and Triantafyllou, M.~S.,
  \enquote{{Oscillating foils of high propulsive efficiency},} \emph{Journal of
  Fluid Mechanics}, Vol. 360, 1998, pp. 41--72.
\newblock \doi{10.1017/S0022112097008392}.

\bibitem[{Carr et~al.(1977)Carr, McAlister, and McCroskey}]{Carr1977}
Carr, L.~W., McAlister, K.~W., and McCroskey, W.~J., \enquote{{Analysis of the
  development of dynamic stall based on oscillating airfoil experiments},}
  Tech. Rep. NASA Technical Note D-8382, NASA, 1977.

\bibitem[{McAlister et~al.(1978{\natexlab{a}})McAlister, Carr, and
  J.}]{McAlister1978}
McAlister, K.~W., Carr, L.~W., and J., M.~W., \enquote{Dynamic Stall
  Experiements on the {NACA} 0012 Airfoil,} Tech. Rep. NASA Technical Paper
  1100, NASA, 1978{\natexlab{a}}.
\newblock \doi{10.1007/BF00575335}.

\bibitem[{McCroskey(1981)}]{McCroskey1981}
McCroskey, W.~J., \enquote{The Phenomenon of Dynamic Stall,} Tech. Rep. NASA
  Technical Memorandum 81264, NASA, 1981.
\newblock \doi{10.1080/6008555886}.

\bibitem[{McAlister et~al.(1978{\natexlab{b}})McAlister, Pucci, McCroskey, and
  Carr}]{McAlister1983}
McAlister, K.~W., Pucci, S.~L., McCroskey, W., and Carr, L.~W., \enquote{An
  Experimental Study of Dynamic Stall on Advanced Airfoil Sections. Volume 2.
  Pressure and Force Data,} Tech. Rep. NASA Technical Memorandum 84245, NASA,
  1978{\natexlab{b}}.

\bibitem[{Seto and Galbraith(1985)}]{Seto1985}
Seto, L.~Y., and Galbraith, R. A.~M., \enquote{The Effect of Pitch Rate on the
  Dynamic Stall of the Effect of Pitch Rate on the Dynamic Stall of a {NACA}
  23012 Aerofoil,} \emph{Eleventh European Rotorcraft Forum}, London, United
  Kingdom, 1985.

\bibitem[{Kaufmann et~al.(2017)Kaufmann, Merz, and Gardner}]{Kaufmann2017}
Kaufmann, K., Merz, C., and Gardner, A., \enquote{Dynamic stall simulations on
  a pitching finite wing,} \emph{Journal of Aircraft}, Vol.~54, No.~4, 2017,
  pp. 1303--1316.
\newblock \doi{10.2514/1.C034020}.

\bibitem[{Visbal and Garmann(2017)}]{Visbal2017}
Visbal, M.~R., and Garmann, D.~J., \enquote{Analysis of Dynamic Stall on a
  Pitching Airfoil Using High-Fidelity Large-Eddy Simulations,} \emph{AIAA
  Journal}, Vol.~56, No.~1, 2017, pp. 0--0.
\newblock \doi{10.2514/1.J056108}.

\bibitem[{Lee and Lua(2018)}]{Lee2018}
Lee, Y.~J., and Lua, K.~B., \enquote{Optimization of Simple and Complex
  Pitching Motions for Flapping Wings in Hover,} \emph{AIAA Journal}, Vol.~56,
  No.~6, 2018, pp. 2466--2470.
\newblock \doi{10.2514/1.B34085}.

\bibitem[{Medina et~al.(2018)Medina, Ol, Greenblatt, M{\"{u}}ller-Vahl, and
  Strangfeld}]{Medina2018}
Medina, A., Ol, M.~V., Greenblatt, D., M{\"{u}}ller-Vahl, H., and Strangfeld,
  C., \enquote{High-Amplitude Surge of a Pitching Airfoil: Complementary Wind-
  and Water-Tunnel Measurements,} \emph{AIAA Journal}, Vol.~56, No.~4, 2018,
  pp. 1--7.
\newblock \doi{10.2514/1.J056408}.

\bibitem[{Zeyghami et~al.(2018)Zeyghami, Zhong, Liu, and Dong}]{Zeyghami2018}
Zeyghami, S., Zhong, Q., Liu, G., and Dong, H., \enquote{Passive Pitching of a
  Flapping Wing in Turning Flight,} \emph{AIAA Journal}, 2018, pp. 1--9.
\newblock \doi{10.2514/1.J056622}.

\bibitem[{{Van Buren} et~al.(2018){Van Buren}, Floryan, and
  Smits}]{VanBuren2018}
{Van Buren}, T., Floryan, D., and Smits, A.~J., \enquote{{Scaling and
  performance of simultaneously heaving and pitching foils},} \emph{AIAA
  Journal}, 2018, pp. 1--12.
\newblock \doi{10.2514/1.J056635}.

\bibitem[{Gursul and Cleaver(2018)}]{Gursul2018}
Gursul, I., and Cleaver, D.~J., \enquote{Plunging Oscillations of Airfoils and
  Wings: Progress, Opportunities, and Challenges,} \emph{AIAA Journal}, 2018,
  pp. 1--18.
\newblock \doi{10.2514/1.J056655}.

\bibitem[{Leishman(1984)}]{Leishman1984}
Leishman, J.~G., \enquote{Contributions to the Experimental Investigation and
  Analysis of Aerofoil Dynamic Stall,} Ph.D. thesis, Department of Aeronautics
  and Fluid Mechanics, University of Glasgow, 1984.

\bibitem[{Freitas(1993)}]{JFEpolicy}
Freitas, C., \enquote{Journal of fluids engineering editorial policy statement
  on the control of numerical accuracy,} \emph{Journal of Fluids Engineering,
  Transactions of the ASME}, Vol. 115, No.~3, 1993, pp. 339--340.
\newblock \doi{10.1115/1.2910144}.

\bibitem[{AIA(2014)}]{AIAAJpolicy}
\enquote{Editorial policy statement on numerical and experimental accuracy,}
  \emph{AIAA Journal}, Vol.~52, No.~1, 2014, p.~16.
\newblock \doi{10.2514/1.J053252}.

\bibitem[{JoA(2010)}]{JoApolicy}
\enquote{Editorial policy statement on numerical and experimental accuracy,}
  \emph{Journal of Aircraft}, Vol.~47, No.~1, 2010, p.~7.
\newblock \doi{10.2514/1.48594}.

\bibitem[{Ghisu et~al.(2015)Ghisu, Puddu, and Cambuli}]{Ghisu_JTS}
Ghisu, T., Puddu, P., and Cambuli, F., \enquote{Numerical Analysis of a {W}ells
  Turbine at Different Non-dimensional Piston Frequencies,} \emph{Journal of
  Thermal Science}, Vol.~24, No.~6, 2015, pp. 535--543.
\newblock \doi{10.1007/s11630-015-0819-6}.

\bibitem[{Ghisu et~al.(2016)Ghisu, Puddu, and Cambuli}]{Ghisu_JFE}
Ghisu, T., Puddu, P., and Cambuli, F., \enquote{Physical Explanation of the
  Hysteresis in {W}ells Turbines: a Critical Reconsideration,} \emph{ASME
  Journal of Fluids Engineering}, Vol. 133, No.~11, 2016.
\newblock \doi{10.1115/1.4033320}.

\bibitem[{Ghisu et~al.(2017)Ghisu, Puddu, and Cambuli}]{Ghisu_JoPaE}
Ghisu, T., Puddu, P., and Cambuli, F., \enquote{A Detailed Analysis of the
  Unsteady Flow within a {W}ells Turbine,} \emph{Proceedings of the Institution
  of Mechanical Engineers Part A Journal of Power and Energy}, Vol. 231, No.~3,
  2017, pp. 197--214.

\bibitem[{Doebelin and Manik(2007)}]{Doebelin}
Doebelin, E.~O., and Manik, D.~N., \emph{Measurement Systems: Application and
  Design}, 5\textsuperscript{th} ed., McGraw-Hill series in Mechanical
  Engineering., Tata McGrawHill Education, New Delhi, 2007.

\end{thebibliography}
\end{document}